# Physically motivated Recursively Embedded Atom Neural Networks: Incorporating Local Completeness and Nonlocality

Yaolong Zhang, Junfan Xia, and Bin Jiang*

*Hefei National Laboratory for Physical Science at the Microscale, Key Laboratory of Surface and Interface Chemistry and Energy Catalysis of Anhui Higher Education Institutes, Department of Chemical Physics, University of Science and Technology of China, Hefei, Anhui 230026, China*

Recent advances in machine-learned interatomic potentials largely benefit from the atomistic representation and locally invariant many-body descriptors. It was however recently argued that including three- (or even four-) body features is incomplete to distinguish specific local structures. Utilizing an embedded density descriptor made by linear combinations of neighboring atomic orbitals and realizing that each orbital coefficient physically depends on its own local environment, we propose a recursively embedded atom neural network model. We formally prove that this model can efficiently incorporate complete many-body correlations without explicitly computing high-order terms. This model not only successfully addresses challenges regarding local completeness and nonlocality in representative systems, but also provides an easy and general way to update local many-body descriptors to have a message-passing form without changing their basic structures.

Over the past years, machine learning has achieved enormous success in many scientific fields, especially in the development of more accurate interatomic potentials based on ab initio data for chemical systems[1], including molecules and reactions[2-7], excited states[8-11], condensed phase materials[12-16], etc. Besides using different machine learning algorithms, these MLIPs mainly differ in their structural descriptors (or features) which should distinguish diverse molecular configurations and be invariant with respect to translation, rotation, and permutation of identical atoms. In small molecular and reactive systems, it is well-known that a global descriptor like permutationally invariant polynomials in terms of interatomic distances[5] of a sufficiently high order, or equivalently fundamental invariants[2], well satisfy both invariance and distinguishability requirements[17]. However, the size of polynomials scales factorially with the number of permutations, preventing their applications in large systems.

On the other hand, most MLIPs for large molecules and materials rely on an atomic decomposition of total energy, namely $E = \sum_{i=1}^{N} E_i$, as first proposed by Behler and Parrinello in their high-dimensional neural network (BPNN) approach[12]. In this representation, each atomic energy is dependent on the corresponding local environment (within a certain cutoff radius) described by a set of locally invariant many-body features between the central and neighboring atoms[18-27]. Due to the high costs of evaluating higher-order terms, these features are typically truncated up to three- or four-body correlations. However, it was recently shown that some local atomic structures in a system as small as $CH_4$ become indistinguishable by the third (or even fourth) order correlations[28]. This would introduce a distortion of the feature space and intrinsically limit the representability of the MLIP[28]. While some approaches[24-26,29] could in principle resolve this atomic structural degeneracy by systematically including higher-order terms, the computational cost would however increase dramatically.

An alternative way to describe an atom-centered environment is to repeatedly convolute feature vectors between every atom and its neighbors by neural networks (NNs) [3], allowing the information progressively passed among the central atom, the neighbors, the neighbors' neighbors, and so on so forth. Such so-called message-passing neural networks (MPNNs)[3,30-32] can learn an increasingly more sophisticated feature-property correlation from the training data. However, it is less clear that how this type of models incorporate many-body correlations by iteratively integrating (mostly) two-body terms[30,31] (and angular terms[33,34]) and whether they can resolve the local structural degeneracy issues discussed in Ref. [28].

In this Letter, to address this challenge, we propose a physically-inspired recursive neural network model that naturally integrates the message-passing concept into a well-defined three-body descriptor. We derive that this model can formulate a complete atomic representation of the local environment without explicitly computing high-order correlations and incorporate some nonlocal interactions beyond the cutoff radius, both validated by numerical tests. Like in conventional MPNN models, however, the nonlocal charge transfer[35] and conjugated effects[1] are not yet included and will not be discussed here.

Let us start with the embedded atom neural network (EANN) model which adopts the atomistic representation of total energy and encodes the information of local environment by the symmetry-invariant embedded density descriptor[20] inspired by the embedded atom method[36].

For simplicity, an embedded density invariant ($\rho_i$) at the position of atom $i$ is given by the square of the linear combination of atomic orbitals of its neighbors,

$$\rho_i = \sum_{l_x,l_y,l_z}^{l_x+l_y+l_z=L} \frac{L!}{l_x!l_y!l_z!} \left[ \sum_{j \neq i}^{N_c} c_j \varphi(\hat{\mathbf{r}}_{ij}) f_c(r_{ij}) \right]^2, \quad (1)$$

where $\hat{\mathbf{r}}_{ij} = \hat{\mathbf{r}}_i - \hat{\mathbf{r}}_j$, with $\hat{\mathbf{r}}_i = (x_i, y_i, z_i)$ and $\hat{\mathbf{r}}_j = (x_j, y_j, z_j)$ being the Cartesian coordinate vectors of the central atom $i$ and a neighbor atom $j$, $r_{ij} = |\hat{\mathbf{r}}_{ij}|$ is the distance between them, $\varphi(\hat{\mathbf{r}}_{ij})$ is the Gaussian-type orbital centered at atom $j$ parameterized by its center ($r_s$), width ($\alpha$), and angular momenta ($L = l_x+l_y+l_z$),

$$\varphi(\hat{\mathbf{r}}_{ij}) = (x_i - x_j)^{l_x}(y_i - y_j)^{l_y}(z_i - z_j)^{l_z} \exp\left[-\alpha(r_{ij} - r_s)^2\right], \quad (2)$$

$f_c(r_{ij})$ is a cutoff function continuously damping the invariant to zero at the cutoff radius ($r_c$), and $N_c$ is the number of atoms within $r_c$. Clearly, $\rho_i$ corresponds to the embedded density contribution from a given type of atomic orbital and expresses two- ($L=0$) and three-body ($L>0$) interactions in a uniform way. This can be seen by explicitly rewriting Eq. (1) in terms of interatomic distances and angles according to the multinomial theorem[20,37],

$$\rho_i = \sum_{j,k \neq i} c_j \exp\left[-\alpha(r_{ij} - r_s)^2\right] f_c(r_{ij}) \\ \times c_k \exp\left[-\alpha(r_{ik} - r_s)^2\right] \times f_c(r_{ik}) r_{ij}^L r_{ik}^L (\cos \theta_{ijk})^L. \quad (3)$$

Indeed, Eq. (1) allows the evaluation of atom-centered three-body terms at a cost of atom-centered two-body ones, resulting in a linear scaling with respect to the number of neighbors. As a result, this EANN model is more efficient than many other descriptor-based MLPs[38], and accurate in predicting energies[20] and tensorial properties[39,40].

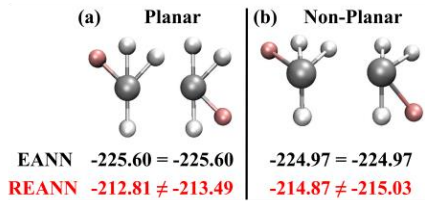

FIG.1 Two representative pairs of CH$_4$ molecules that have the same set of distances and angles between central C atom (silver) and neighboring H atoms (white for identical ones and light red for the different one), for which the EANN (REANN) atomic energies (in eV.) for C are identical (distinct).

In the CH$_4$ example, the C-centered embedded density invariants and corresponding atomic energies are essentially identical, when two C-centered structures of CH$_4$ have the same list of distances and angles, as displayed in FIG. 1. This problem intrinsically exists in other three-body (or lower-order) atomic descriptors [18,19,22,23]. It can be seen from Eq. (3) as orbital coefficients are fixed after training (like NNs' parameters) so that $\rho_i$ are determined by these distances and angles only. However, considering the linear combination of atomic orbitals in Eq. (1) as an analog of a molecular orbital, it is a matter of fact in quantum chemistry that $c_j$ should virtually vary with the molecular configuration. One simplest way to cast this physical concept into the descriptor is to make $c_j$ itself a function of the $j$th atom's neighbor environment behaving like the atomic energy. In this scenario, orbital coefficients of the four H atoms in the two CH$_4$ molecules can be different since their respective H-centered environments are different. This leads to nonequivalent C-centered embedded density invariants and atomic energies for the two indistinguishable atomic structures by three-body correlations in Fig. 1. Importantly, atomic orbitals in the vicinity of atom $j$ have been calculated for obtaining the atomic energy ($E_j$), thus need not be recalculated to evaluate the environment-dependent $c_j$.

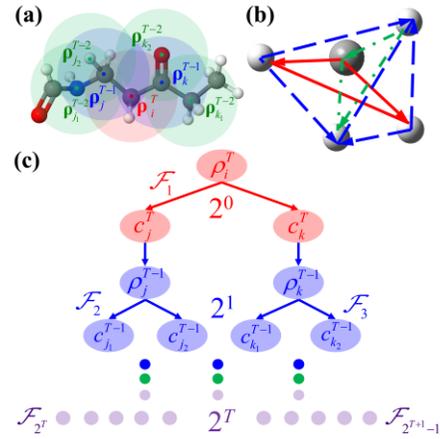

FIG. 2 (1) Schematic diagram of the REANN model showing how the density descriptor is recursively embedded. (b) An example of CH$_4$ showing how the C-centered complete five-body correlation is achieved by twice iteration, where the path going through all atoms correspond to the product of $\mathcal{F}$ functions and arrows point from the central atom to neighbor atoms. (c) An illustration that how the number of three-body terms ($F$) increases in each iteration ($2^T$). Different colors correspond to different iteration times, namely $T=0$ (red), $T=1$ (blue), $T=2$ (green).

Apparently, the orbital coefficient can be recursively embedded in this way whenever necessary and a generalized expression is,

$$c_j^t = g_j^{t-1}\left(\mathbf{\rho}_j^{t-1}\left(\mathbf{c}_j^{t-1}, \mathbf{r}_j^{t-1}\right)\right), \quad (4)$$

where $\mathbf{c}_j^{t-1}$ and $\mathbf{r}_j^{t-1}$ are the collections of orbital coefficients and atomic positions in the neighborhood of the central atom $j$ in the ($t$-1)th iteration, $\mathbf{\rho}_j^{t-1}$ is the corresponding embedded density feature vector, $g_j^{t-1}$ is an atomic NN mapping $\mathbf{\rho}_j^{t-1}$ to $c_j^t$, namely the orbital coefficient of atom $j$ as a neighbor of other atoms in the $t$th iteration. This procedure is schematically displayed in Fig. 2a. One may immediately realize that this recursively EANN (REANN) model has an effective message-passing form[1], except that here the orbital coefficients, rather than the whole feature vectors, iteratively pass the environmental information between an atom and its neighbors. This is an intriguing result that links up, perhaps for the first time, the local many-body descriptors and the less physically intuitive message-passing features.

Next we turn to discuss how higher-order correlations are incorporated in this recursion, an issue rarely discussed in previous studies on MPNNs. Supposing that the iteration undergoes $T$ times ($T>0$), it is convenient to use a simplified version of Eq. (3),

$$\rho_i^T = \sum_{j,k} c_j^T c_k^T \mathcal{F}(r_{ij}, r_{ik}, r_{jk}), \quad (5)$$

where the orbital coefficients are now $T$-dependent ($c_j^T$ and $c_k^T$) and $\mathcal{F}(r_{ij}, r_{ik}, r_{jk})$ represents a generalized three-body correlation term collecting these functions in Eq. (3). Substituting Eq. (4) into Eq. (5) and assuming no hidden layer in $g_j^{T-1}$ (i.e. a linear function), we have

$$\rho_i^T = \sum_{j,k} \mathcal{F}(r_{ij}, r_{ik}, r_{jk}) \sum_{n_1=1}^{N_\rho} w_{n_1} \rho_j^{T-1, n_1} \sum_{n_2=1}^{N_\rho} w_{n_2} \rho_k^{T-1, n_2}, \quad (6)$$

where $w_{n_1}$ and $w_{n_2}$ are linear weights of the corresponding features, $N_\rho$ is the number of embedded density invariants. Note that using a nonlinear $g_j^{T-1}$ here would not alter our conclusion but will complicate this equation. We then substitute Eq. (5) in the ($T$-1)th iteration back to Eq. (6) and reorder the summations,

$$\rho_i^T = \sum_{j,k}\sum_{n_1,n_2}^{N_\rho} w_{n_1} w_{n_2} \sum_{j_1,j_2}\sum_{k_1,k_2} c_{j_1}^{T-1,n_1} c_{j_2}^{T-1,n_1} c_{k_1}^{T-1,n_2} c_{k_2}^{T-1,n_2} \mathcal{F}(r_{ij}, r_{ik}, r_{jk})$$
$$\times \mathcal{F}(r_{jj_1}, r_{jj_2}, r_{j_1 j_2}) \mathcal{F}(r_{kk_1}, r_{kk_2}, r_{k_1 k_2}), \quad (7)$$

As orbital coefficients are expanded, the number of three-body functions doubles in each iteration till the last environment-independent ones, as illustrated in Fig. 2c. This will make $\rho_i^T$ eventually the sum of products of $(2^{T+1}-1)$ three-body $\mathcal{F}$ functions after $T$ iterations, which can be generalized as,

$$\rho_i^T = \sum_m \eta_m \prod_{(i,j,k)}^{2^{T+1}-1} \mathcal{F}(r_{ij}, r_{ik}, r_{jk}), \quad (8)$$

where $m$ collects all indexes of the summation, $\eta_m$ is the collection of all weights and orbital coefficients, and $i$, $j$, $k$ span over all atomic indexes involved. According to above discussion, $\rho_i^T$ will contain at least one highest-order correlation term involving $3(2^{T+1}-1)$ non-redundant interatomic distances in the neighborhood of atom $i$ with sufficient neighbors, along with some lower-order terms due to repeated interatomic distances. Regarding atoms as nodes and interatomic distances as edges, the highest-order correlation term can be viewed as an analog of the Eulerian path in graph theory (a path in a finite graph passing every edge just once), except that in our case this path can pass the same edge more than once. Fig. 2b illustrates such a path walking through all edges in $CH_4$ after two iterations. Examples for lower-order correlations are provided in the Supplemental Material (SM)[41].

By definition, a complete many-body descriptor has to correlate all atoms in the system[46]. This implies that $\rho_i^T$ will involve a complete correlation of an atom-centered environment, if $3(2^{T+1}-1) \geq N_c(N_c-1)/2$. The minimum number of iterations to warrant this is thus given by $T_{min} = [\log_2\{(N_c(N_c-1))/6+1\}]-1$, where [] rounds up the value to its nearest integer. Recall that the cost of each iteration scales linearly with $N_c$ and atomic orbitals need be calculated only once. This is a striking finding that the complete atomic representation can be achieved with $\sim O(\log_2 N_c)$ complexity, instead of the exponential scaling with the body-order when explicitly computing high-order correlations[29]. Our approach will be increasingly more favorable as $N_c$ increases.

Similarly, this analysis can also estimate the required number of interaction blocks (or the time of message passed) in other MPNN models, which was often empirically specified without a guidance. This number has to be greater than $N_c(N_c-1)/2$, theoretically, if only two-body features were recursively embedded (e.g. in SchNet[30]), because each iteration now introduces only one more interatomic distance towards the higher-body correlation. It is even worse that using radial functions alone actually does not warrant the local completeness, because atoms with distances greater than $r_c$ cannot be correlated in any way. Examples are given in the SM. It is also found in other more recent MPNN models that including angular information in the feature update is beneficial[33,34], consistent with our derivation. Note that our practical implementation remains based on Eqs. (2) and (4) for numerical efficiency.

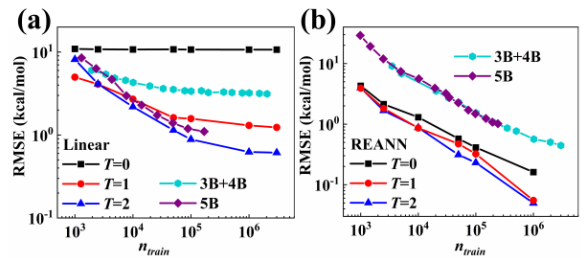

FIG. 3 (a) Comparison of the RMSEs for energies of random $CH_4$ configurations of linear fits in Refs. [28] (with 3B+4B correlations) and [29] (with 5B correlation), and that with the recursively embedded density descriptor ($T$=0, 1, 2) (b) Similar to (a) but all results are now based on nonlinear NN fits.

To validate our derivation numerically, we use the $CH_4$ dataset provided by Ceriotti and coworkers as a stringent test[28]. This dataset includes ~7.7 million configurations with randomly distributed atoms excluding structures with too close contacts. Due to the existence of near degenerate manifolds and many unphysical configurations with energies up to 70 eV, this dataset has been claimed to be the best touchstone of the representability and completeness of the descriptor. Since there are only five atoms in $CH_4$, we estimate that many-body correlations become complete at $T_{min} = 2$. We have optimized $r_s$, $\alpha$, and $c_j$ together with all NN parameters, as readily implemented in PyTorch[42], yielding an end-to-end deep learning framework. To demonstrate the performance of the features themselves, we also train linear models by removing all hidden layers of NNs (for both orbital coefficients and atomic energies). Details of training are given in the SM.

Fig. 3a compares the test root-mean-square-errors (RMSEs) of various linear models as a function of the number of training configurations ($n_{train}$). The learning curve of $T=0$ (including three-body correlations only) exhibits a clear saturation with respect to $n_{train}$, which is fully consistent with the result of Ref. [28] using three-body power spectrum features. Recursively expanding orbital coefficients steepens the learning curve and reduces the error significantly. The result with a single iteration ($T=1$) obviously outperforms that from Ref. [28] obtained with the mix of three- and four-body (3B+4B) correlations. With two iterations ($T=2$), which are supposed to offer a complete correlation, we observe a saturated error of ~0.6 kcal/mol with $10^6$ points. This is in good agreement with that of Nigam et al.[29] who used an iterative contraction algorithm to select up to five-body (5B) invariants (the highest-body correlation for $CH_4$). These results clearly indicate the local completeness of our recursively embedded density descriptor.

Incorporating the nonlinearity of NNs substantially increases the flexibilities of all models. As shown in Fig 3b, 3B+4B correlations in Ref. [28] trained with 3 million points led to an RMSE of ~0.5 kcal/mol. Impressively, our EANN model ($T=0$) gives a much lower learning curve, exhibiting its superior performance despite its three-body nature. The lower error may be due to the deeper NNs used in our EANN model, but one shall note that much fewer invariants are used as the input (45) here than that (2000) in Ref. [28]. The model accuracy increases with $T$, although the improvement from $T=1$ to $T=2$ is less significant than that from $T=0$ to $T=1$. This is consistent with the fact that $T=1$ already includes eight interatomic distances of $CH_4$ (see Fig. 2b) that are close to complete (10 distances in total). The learning curve more or less converges at $T=2$, whose errors are one order of magnitude smaller than those with 3B+4B features[28], and those with the contracted 5B features[29]. We find that other MPNN models[30,33,34] also perform better than the purely local descriptor-based model[28] (detailed in the SM). This provides more convincing evidence that iterative message-passing can include more complete correlations to better represent the atomic environment.

An additional advantageous feature of the REANN model is its effective description of some nonlocal effects. This is because the correlations between atoms inside and outside the cutoff sphere have been implicitly encoded when iteratively updating orbital coefficients, as illustrated in Fig. 2a. We demonstrate this in bulk water, an important benchmark to demonstrate the power of MLIPs. We first use a dataset with 1593 structures of 64 water molecules computed by Cheng et al. [45] for developing a BPNN potential[12]. The cutoff radius of BPNN potential was set long enough ($r_c$=6.2 Å) to describe the strong hydrogen bond interactions. With an optimal selection of symmetry functions (3B features), the reported RMSEs of the BPNN potential are comparable to those of the EANN model[38] with the same $r_c$, as listed in Table 1. Impressively, our REANN model ($T$=3) greatly outperforms these two purely local descriptor-based counterparts, leading to a smaller error of force with merely half of the cutoff radius ($r_c$=3 Å). Apparently, this cutoff only incorporates the interactions between a water molecule and some nearest neighbors, but the second neighboring shell is implicitly correlated by the message-passing way of updating orbital coefficients. The performance of the REANN model further improves with the increasing $r_c$ and saturates at $r_c \approx 5.5$ Å, yielding less than half of the RMSEs of the BPNN potential. It is worth noting that the SchNet model performs less well in this condensed phase system, due presumably to that more interatomic distances are greater than $r_c$ and corresponding atoms cannot be correlated by two-body terms. To avoid any data bias, we simply test another dataset of water trained by Zhang et al. using the deep potential molecular dynamics (DPMD) method[14] with $r_c$=6.0 Å. Our REANN model ($T$=2) with $r_c$=4.5 Å gives RMSEs of 0.2 meV/atom (energies) and 15.9 meV/Å (forces), again less than half of the reported values in Ref. [14]. These results suggest that the REANN model captures nonlocal interactions more efficiently than simply extending the effective cutoff radius in complex systems.

Table 1: Test RMSEs of energies (meV/atom) and forces (meV/Å) for bulk water using the dataset in Ref. [45].

| Model | REANN ($T$=3) | | | | EANN* | BPNN |
|---|---|---|---|---|---|---|
| $r_c$ (Å) | 3.0 | 3.5 | 4.5 | 5.5 | 6.2 | 6.2 | 6.2 |
| Energy | 2.8 | 1.5 | 1.1 | 0.9 | 0.8 | 2.1 | 2.3 |
| Force | 104.4 | 73.1 | 58.0 | 51.1 | 53.2 | 129.0 | 120 |

*Values taken from Ref. [38].

Summarizing, we make a physical adaption of the local descriptor-based EANN model to generate the REANN model and reveal its connection with other less physically-intuitive MPNNs often inspired from graph neural networks in computer science. We formally derive that how the many-body correlations are introduced by iteratively passing messages (updating orbital coefficients here) and prove that this is a more efficient way to achieve a complete description of the local environment, without explicitly computing high-order features. Numerical tests demonstrate the local completeness and nonlocality of this new model, warranting its superior accuracy among existing ML models. Our strategy can be easily adapted to improve other sophisticated many-body descriptors without changing their basic structures, for example, by making atomic weights of the weighted atom-centered symmetry functions variable with its local environment[23] or adding such learnable coefficients to the DPMD descriptors[14]. We believe this will open a new window for developing more accurate and efficient ML models of more complicated physical systems.

**Acknowledgement:** This work is supported by National Key R&D Program of China (2017YFA0303500), National Natural Science Foundation of China (22073089 and 22033007), Anhui Initiative in Quantum Information Technologies (AHY090200), and The Fundamental Research Funds for Central Universities (WK2060000017). Calculations have been done on the Supercomputing Center of USTC. We thank Prof. Michele Ceriotti and Dr. Sergey Pozdnyakov for kindly explaining their dataset for $CH_4$.

*bjiangch@ustc.edu.cn

# Supplemental Material

# Physically-motivated Recursively Embedded Atom Neural Networks:

# Incorporating Local Completeness and Nonlocality

Yaolong Zhang, Junfan Xia, and Bin Jiang[*]

*Hefei National Laboratory for Physical Science at the Microscale, Key Laboratory of Surface and Interface Chemistry and Energy Catalysis of Anhui Higher Education Institutes, Department of Chemical Physics, University of Science and Technology of China, Hefei, Anhui 230026, China*

*: corresponding author: bjiangch@ustc.edu.cn

# I. Training Details

## A. Neural network implementation

The proposed recursively embedded atom neural networks (REANN) package is implemented in PyTorch[42], which allows for highly efficient optimization of all parameters. In practice, all hyperparameters including $\alpha$, $r_s$ and orbital coefficients $c_j$ were optimized along with the weights and biases of neural networks (NNs) to minimize the cost function using mini-batch gradients algorithm. The AdamW optimizer[43] with default setup was employed and the "ReduceLROnPlateau" learning rate scheduler provided by PyTorch[42] was adopted to decay the learning rate from $10^{-3}$ to $10^{-5}$ with a factor 0.6. Both energies and forces were included in the cost function with a dynamic weight of the atomic force vectors $\beta$ which decreased from an initial value ($\beta_{init}=5$) to a final value ($\beta_{end}=1$). We used the softplus activation function in all cases. Other details about the NN architectures and parameters of embedded density descriptor for each system are listed in Table S1.

## A. CH$_4$

The CH$_4$ dataset was provided by Ceriotti and coworkers[28]. This dataset was designed for checking the distinguishability of the structural descriptors, containing many near-degenerate configurations with distinct energies and unphysical configurations with energies up to 70 eV. The total energies and forces were computed at the GGA-DFT level. Only those structures for which the self-consistent calculations converged in a reasonable number of iterations (5 times more than the default value).

This procedure more or less excluded those points that converged to wrong energies that would introduce numerical noises to the dataset. Due to the structural degeneracy, it was argued that the machine learning model with lower-order body features would become less accurate than that with higher-order body features, given sufficiently rich data[28]. In addition, the prediction accuracy would saturate with increasing training set size at a sufficiently large number of training configurations.

The learning curves obtained in FIG. 3 were generated by varying the size of training set from $10^3$ to $3\times10^6$ randomly selected points. The reported root-mean-square-errors (RMSEs) in this work were evaluated with the same test set (with 80000 randomly chosen structures) used in Ref. [28]. In the $CH_4$ system, two residual NN blocks ($n_{block}$=2) were employed and a dropout algorithm that randomly zeroes some of the neurons in each layer with a probability $p$ (0~0.3) was adopted to prevent overfitting[44].

**B. Bulk water**

Two independent datasets of bulk water reported by Cheng *et al.*[45] and Zhang *et al.*[14], referred as water-I and water-II, respectively, were taken to demonstrate the performance of the new REANN model. We followed the training setups in these two studies. The first dataset contains 1593 structures calculated at revPBE0-D3 level covering a large configuration space, with 64 water molecules in each configuration in periodic condition. The gross data were randomly separated with a ratio of 80% and 20% for training and testing, respectively, as done in Ref. [45]. The second dataset of Zhang *et al.* consists of ~$10^5$ structures extracted from path integral ab initio molecular dynamics trajectories at 300 K, computed by the hybrid version of PBE0 and Tkatchenko-Scheffler functional. Ninety percent of data points were randomly chosen for training, and the remainder were used for testing. We kept the size of NNs and the number of invariant features similar to those reported in original publications to allow a fair comparison, which can be found in Table S1.

## II. Illustrations of high order correlations

In the REANN framework, we argue in the main text that at least one highest-order correlation term will appear in $\rho_i^T$, after a sufficiently number of recursive embedding of the orbital coefficients. We show an example for $CH_4$ in FIG. 2b, where a five-body (5B) correlation (complete for $CH_4$) involving fully-interconnected atoms is achieved with twice iteration. $\rho_i^T$ also contains lower-order terms for which some interatomic distances repeatedly appear in the many-body functions. In FIG. S1, we further show exemplary four-body (4B), three-body (3B) and two-body (2B) interactions, corresponding to the following many-body functions in the final expression of $\rho_i^{T=2}$,

$$\Gamma_{4B} = \mathcal{F}(r_{C,H1}, r_{C,H2}, r_{H1,H2})\mathcal{F}(r_{H1,C}, r_{H1,H3}, r_{C,H3})\mathcal{F}(r_{H2,C}, r_{H2,H3}, r_{C,H3})$$
$$\times \mathcal{F}(r_{C,H1}, r_{C,H3}, r_{H1,H3})\mathcal{F}(r_{H3,C}, r_{H3,H1}, r_{H1,H3})\mathcal{F}(r_{C,H1}, r_{C,H3}, r_{H1,H3})\mathcal{F}(r_{H3,C}, r_{H3,H2}, r_{H2,H3})$$
, (1)

$$\Gamma_{3B} = \mathcal{F}(r_{C,H1}, r_{C,H2}, r_{H1,H2})\mathcal{F}(r_{H1,C}, r_{H1,H2}, r_{C,H2})\mathcal{F}(r_{H2,C}, r_{H2,H1}, r_{C,H1})$$
$$\times \mathcal{F}(r_{C,H1}, r_{C,H2}, r_{H1,H2})\mathcal{F}(r_{H2,C}, r_{H2,H1}, r_{C,H1})\mathcal{F}(r_{C,H1}, r_{C,H2}, r_{H1,H2})\mathcal{F}(r_{H1,C}, r_{H1,H2}, r_{H2,C})$$,

(2)

$$\Gamma_{2B} = \mathcal{F}(r_{C,H1})\mathcal{F}(r_{H1,C})\mathcal{F}(r_{C,H1}). \tag{3}$$

Note that the last two-body function is produced by $L=0$, where the initial embedded density descriptor contains two-body terms only.

**III. Incompleteness of two-body feature-based message passing neural networks**

Our derivation in the main text explains how the many-body interactions are iteratively encoded into the message-passing type of neural network (MPNN) models. In commonly-used MPNNs such as SchNet[30] and PhysNet[31], only two-body features (radial functions) were recursively embedded. The key feature of MPNNs is that each interaction block will only pass the interactions within an atom-centered cutoff sphere ($r_c$) to the next block. However, two-body functions between the center atom and neighbor atoms cannot correlate two neighbor atoms themselves. As a result, the correlations between neighbor atoms with distances larger than $r_c$ cannot be involved no matter how many messages are passed, thus preventing a complete representation for such atomic structures. FIG. S2 illustrates an atomic structure of $CH_4$ in which the four neighboring H atoms are placed at the edge of the cutoff sphere. While this is a rather unphysical configuration for $CH_4$, similar configurations with two (or more) neighbor atoms in the local environment distant from each other beyond $r_c$ are quite common in periodic systems. After several message-passing iterations, although these C-H correlations have been repeatedly incorporated, no additional correlations between any two H atoms with distances greater than $r_c$ are established. This is completely different from the description of REANN, as shown in FIG. S2b, in which the interactions between two neighbor atoms with distances larger than $r_c$ can be introduced by the angular terms.

In order to illustrate this in a more straightforward way, we present numerical tests of two representative $CH_4$ configurations which cannot be distinguished by SchNet. These structures are shown in FIG. S3, along with the corresponding energies predicted by SchNet and REANN models. We elongate the four C-H bonds from the $CH_4$ equilibrium to 4.5 Å to obtain structure (a) so that all H-H distances exceed the cutoff radius ($r_c$=6.5 Å here). Rotating one of the C-H bonds (marked in light red) to some extent and keeping other atomic positions unchanged results in structure (b), whose energy should be different from that of (a). However, since SchNet relies on two-body interactions only, all local features of the C- and H-centers are equivalent for the two configurations. Although their H-H distances are different, they are all greater than the cutoff and thus not able to contribute to local features in the SchNet framework. This leads to exactly the same energies for the two configurations. By contrast, the REANN model captures the differences in the H-H distances implicitly by three-body interaction and predicts different energies for them. This example provides unambiguous evidence to support our argument that three-body features are the minimum requirement for MPNNs to achieve local completeness in an arbitrary configuration. Higher-order interactions can be of course used in the same way and will help reduce the required number of iterations to converge the local completeness. One shall however note that computing higher-order interactions itself is much more expensive than computing lower-order ones. How to optimally combining many-body features with message-passing to achieve the best accuracy and efficiency for a given system is still an open question, but this is beyond the scope of this work.

It is worth noting that the example configurations are not absolutely unreasonable and could be relevant to the multi-body fragmentation of $CH_4$, though they do not exist in the $CH_4$ dataset from Ref. [28] so that the model predicted energies seem unrealistic. As argued in Ref. [28], the existence of manifolds of degenerate structures introduces a distortion of the feature space, which hinders the ability to perform regression. Its influence on the model accuracy is therefore more implicit. In spite of this intrinsic issue, we are not saying that the practical SchNet-based interaction potentials will be problematic. Indeed, this issue will become less severe as the cutoff radius increases. One can always lift the cutoff radius to cover as many atoms as possible in molecular systems, but it is difficult to do so in periodic systems, as discussed in the next section.

**IV. Comparison with other message passing neural networks**

In order to demonstrate the general performance of MPNN models, we perform additional calculations using the freely available codes of SchNet[30] (a widely-used two-body-feature based MPNN model), DimeNet[34] and Cormorant[33] (currently published MPNN models with angular features). It should be noted that the DimeNet and Cormorant programs neither include atomic force learning nor periodic boundary conditions, which can be applied only to the $CH_4$ system, yet not the bulk water system. Because of their diverse architectures and various factors affecting the model accuracy, it is difficult to evaluate different models on equal-footing. We have not aimed to optimize each model and tried to use the default setting of each program that has proven to work well for benchmark datasets (*e.g.* QM9 and MD17). Consequently, this comparison is more like a proof-of-concept. The trained models are freely available in https://github.com/zhangylch/Physically-motivated-Recursively-Embedded-Atom-Neural-Networks. Note that the number of message-passing iteration $T$ follows the definition in this work, namely $T$=0 corresponds to the regular atomic descriptor with no message-passing.

For the CH$_4$ data set, a cutoff radius ($r_c$=6.5 Å) is employed for all models except Cormorant which uses a learnable cutoff[33]. All message-passing based models are more accurate than the model reported in Ref. [29]. with contracted 5B features. This more generally supports our conclusion that higher-order correlations can be effectively incorporated by message-passing of low-order ones. While all based on message-passing of three-body interactions, REANN ($T$=2) performs better than DimeNet ($T$=3) and Cormorant ($T$=2) models, probably because the latter two models learn energies only. DimeNet becomes increasingly more accurate than Cormorant in the data-sufficient region. Importantly, since the H-H distance is always smaller than 6.5 Å in the dataset, SchNet is capable of incorporating H-H correlations via message-passing. As a result, the SchNet model performs comparably well as the REANN model in this case, despite requiring more iterations ($T$=5).

On the other hand, for the periodic dataset of Cheng *et al.*[45] for bulk water, the interatomic distances are more likely greater than the cutoff radius. In this case, we found that the SchNet model with 64 radial functions and 64 neurons in each layer perform better than that by default. A detailed comparison between SchNet with different $T$ and cutoff radius and REANN is given in the Table S2. With the same $r_c$=5.5 Å, the SchNet model ($T$=5, by default) results in test RMSEs of 2.3 meV/atom (energy) and 143 meV/Å (force), which are comparable to those of the BPNN[45] and EANN[38] models, while almost three times as large as those of the REANN ($T$=3) model. Reducing the number of message-passing iterations to $T$=3 for SchNet leads to the absence of some higher-order correlations, which increases errors to 3.1 meV(energy) and 170 meV/ Å (force). On the other hand, shortening $r_c$ to 3.0 Å greatly increases the errors of the SchNet model to 4.3 meV/atom (energy) and 205 meV/ Å (force). This is consistent with our earlier discussion that a smaller cutoff radius may introduce more degenerate configurations indistinguishable by the two-body-interaction based model, hindering its representability. All these results in fact further strengthen our conclusion in this work.

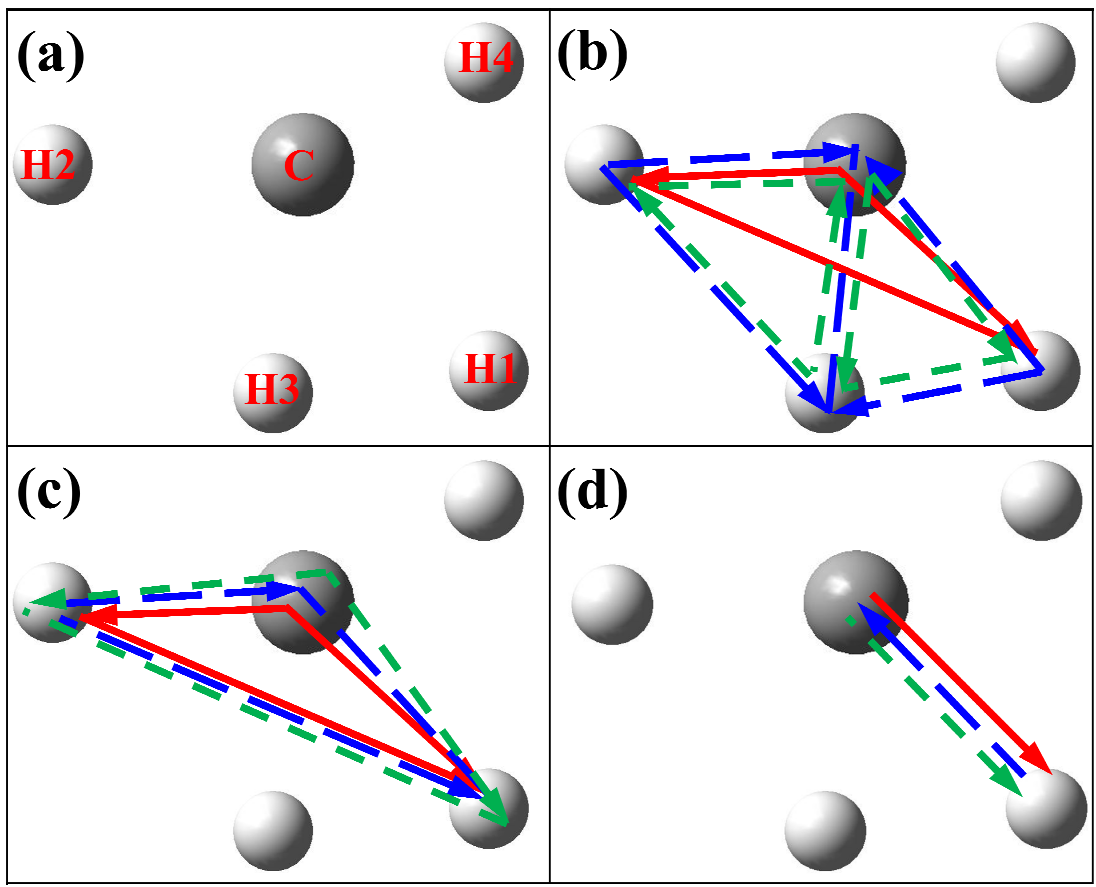

FIG S1: (a) A representative configuration of CH$_4$ with atom labeling, for which (b) four-body, (c) three-body, and (d) two-body correlations of the C-centered atomic structure are iteratively incorporated in the REANN framework. Each path connecting correlated atoms corresponds to a product of $\mathcal{F}$ functions. Different colors and types of lines correspond to different iterations, namely $T=0$ (red, solid), $T=1$ (blue, long dashed) and $T=2$ (green, medium dashed). Arrows point from the central atom to the neighbor atom in each iteration.

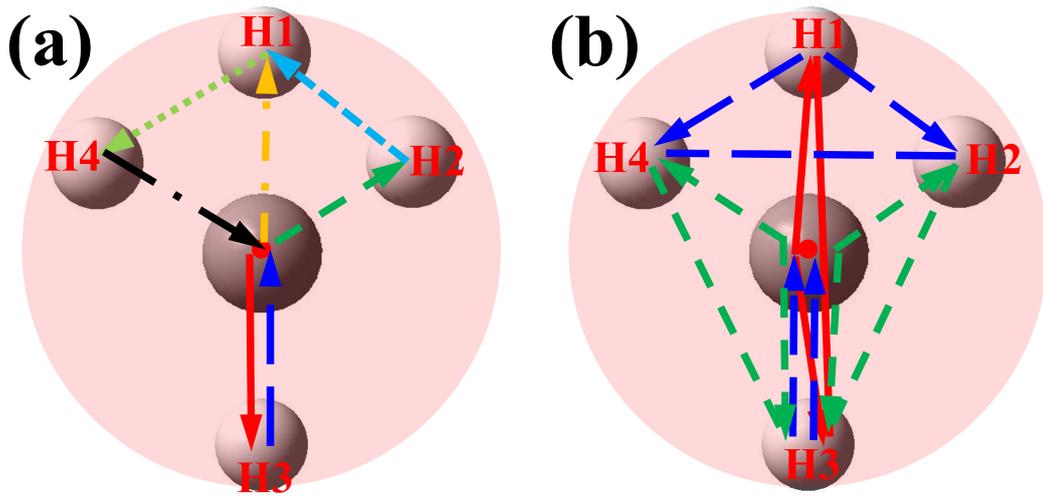

FIG S2: Illustrations of many-body correlations evolved with the "message" passed in (a) two-body-feature-based MPNN models and (b) the three-body-feature-based REANN model. Each path connecting correlated atoms corresponds to a product of $\mathcal{F}$ functions. Different colors and types of lines represent different iterations, namely $T=0$ (red, solid), $T=1$ (blue, long dash), $T=2$ (green, medium dash), $T=3$ (light blue, short dash), $T=4$ (light green, dot), $T=5$ (black, long dash dot), $T=6$ (orange, medium dash dot). Arrows point from the central atom to the neighbor atom in each iteration. Here, the distances between H1-H3, H2-H3, H4-H3 and H2-H4 are larger than the cutoff radius, and the correlations among them can not be established by two-body-feature-based MPNN models.

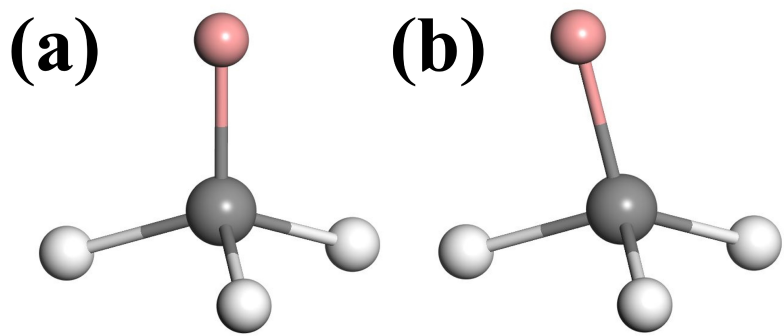

SchNet  -1235.29  =  -1235.29
REANN  -1106.69  ≠  -1107.98

FIG S3: Two representative CH₄ structures (a) and (b) with the same C-H distances (4.5 Å) but different H-C-H angles. The single different H atom is marked in light red in the two structures. Since all H-H distances are greater than the cutoff radius (6.5 Å), the correlations among them are not included in SchNet, which predicts the same energy for the two configurations. The REANN model does distinguish them with different energies. Energies are in eV and the minimum energy in the CH₄ dataset is -1098.76512 eV.

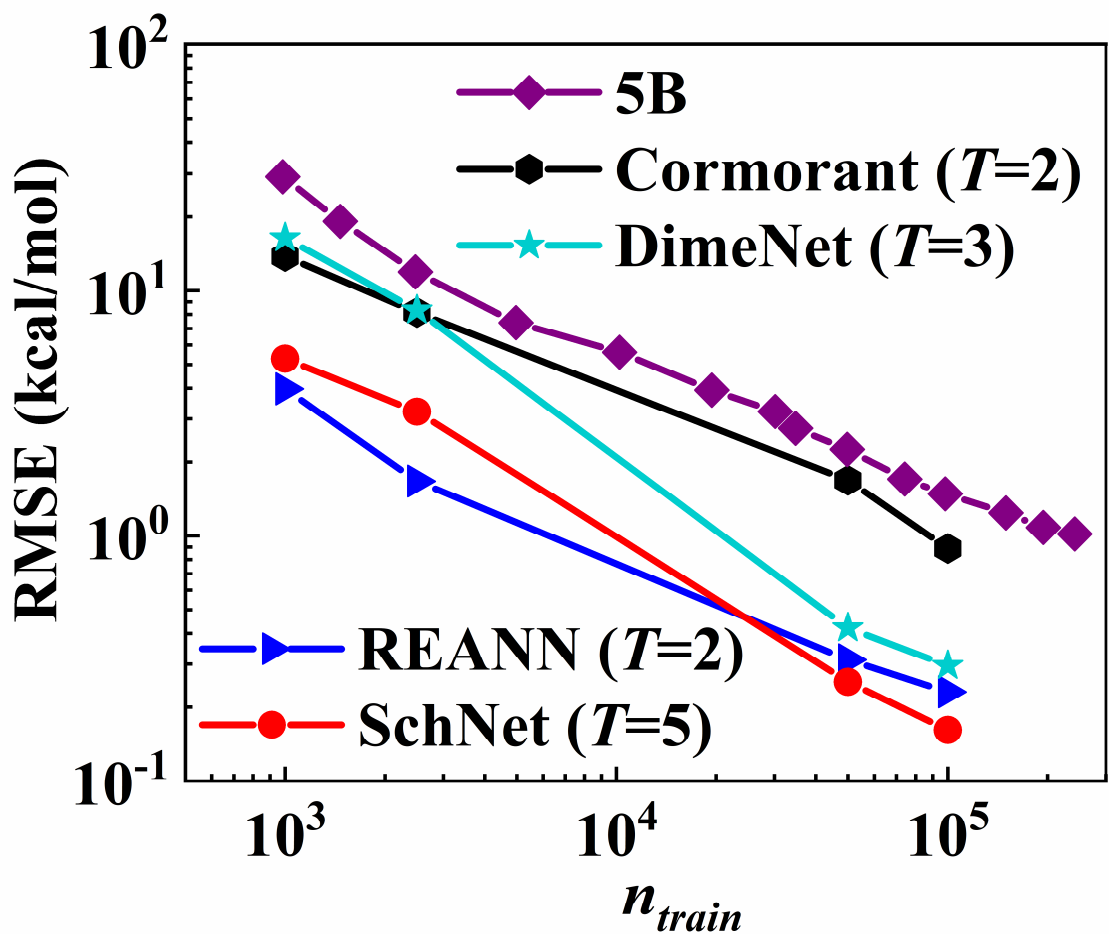

FIG S4: Comparison of the test RMSEs for energies of a random dataset of $CH_4$ predicted by SchNet, Cormorant, DimeNet, REANN models and a neural network model reported in Ref. [29] with contracted five-body features. The number of message-passing iterations is given in the parenthesis for each model and $T=0$ corresponds to no massage-passing.

Table S1: NN structures (denoted by the number of neurons for each hidden layer), cutoff radii ($r_c$), number of features $\rho_i^T$ ($N_\rho$) used in the training processes.

| System | NN structure[a] | $r_c$ | $N_\rho$ | NN structure[b] |
|---|---|---|---|---|
| $CH_4$[c] | 512×512×512×512×512 | 6.5 | 45 | 256×256 |
| $CH_4$ (Linear) | \ | 6.5 | 120 | \ |
| Water-I | 32×16 | 6.2 | 33 | 16×16 |
| Water-II | 512×256×128×64 | 6.0 | 60 | 64×32 |

[a]Atomic NN structure for atomic energies.

[b]Atomic NN structure for orbital coefficients.

[c]For $CH_4$, two residual NN blocks ($n_{block}$=2) were employed. Each residual NN structure is given here.

Table S2: Test RMSEs of energies (meV/atom) and forces (meV/Å) for bulk water using the dataset in Ref. [45].

| Model | REANN ($T=3$) | | SchNet ($T=5$) | | SchNet ($T=3$) | EANN[*] | BPNN |
|---|---|---|---|---|---|---|---|
| $r_c$ (Å) | 3.0 | 5.5 | 3.0 | 5.5 | 5.5 | 6.2 | 6.2 |
| Energy | 2.8 | 0.9 | 4.3 | 2.3 | 3.1 | 2.1 | 2.3 |
| Force | 104.4 | 51.1 | 204.6 | 143.0 | 170 | 129.0 | 120 |

[*]Values taken from Ref. [38].